\documentclass[prb,twocolumn,showpacs,amssymb,amsmath,floats]{revtex4}

\usepackage{epsf}
\begin{document}

\def\be{\begin{equation}}
\def\ee{\end{equation}}
\def\bea{\begin{eqnarray}}
\def\eea{\end{eqnarray}}
\def\br{{\bf r}}
\def\bv{{\bf v}}
\def\bk{{\bf k}}
\def\bn{{\bf n}}
\newcommand{\corr}[1]{\langle #1\rangle}
\newcommand{\Corr}[1]{\left\langle #1\right\rangle}
\def\eps{\varepsilon}
\def\tr{\mathop{\rm tr}}

\title{Vortex viscosity in the moderately clean
limit of layered superconductors}

\author{M. A. Skvortsov$^1$, D. A. Ivanov$^2$, and G. Blatter$^2$}

\affiliation{$^1$L. D. Landau Institute for Theoretical Physics,
Moscow
117940, Russia\\
$^2$Institut f\"ur Theoretische Physik, ETH-H\"onggerberg, CH-8093
Z\"urich, Switzerland}

\begin{abstract}
We present a microscopic calculation of the energy dissipation in
the core of a vortex moving in a two-dimensional or layered
superconductor in the moderately clean regime. In this regime, the
quasiclassical Bardeen--Stephen result remains valid in spite of
the strong correlations between the energy levels. We find that
the quasiclassical expression applies both in the limit of fast
vortex motion (with transitions between smeared levels) and in the
limit of slow vortex motion (with nearly adiabatic dynamics).
This finding can be related to the similar result known for
the unitary random-matrix model.
\end{abstract}


\maketitle

\section{Introduction}

At low temperature the energy dissipation in the vortex cores is
the main source of resistivity in the mixed state of type-II
superconductors\cite{Larkin-Ovchinnikov}. If a supercurrent flows
through a superconductor, it exerts a force on the vortices.
Unless pinned by impurities or inhomogeneities, the vortices are
brought into motion, which in turn leads to dissipation. Depending
on the level of disorder, the vortices may move at different
angles with respect to the direction of the supercurrent. At weak
disorder, the vortices move together with the supercurrent
(``ballistic limit'', with the Hall angle close to $\pi/2$).
At sufficiently strong disorder, the vortex motion is directed
perpendicular to the supercurrent (``dissipative limit'', small
Hall angle). Both limits are well understood within the
quasiclassical description
\cite{GK73,Bardeen-Sherman,LO76,Kopnin-Kravtsov,Kopnin94,Kopnin-Lopatin}.
A simplified approach describing the dissipative limit goes back
to the theory of Bardeen and Stephen treating the vortex as a
region of normal phase inside a superconductor
\cite{Bardeen-Stephen}. In spite of neglecting the structure of
the quasiparticle excitations in the vortex core \cite{CDGM}, the
Bardeen--Stephen theory gives the same result (up to a numerical
factor) as the accurate quasiclassical calculation
\cite{GK73,Bardeen-Sherman,LO76,Kopnin-Kravtsov,Kopnin94,Kopnin-Lopatin}.

It has been recently suggested that the microscopic structure of
the core excitations may play a much more prominent role in that
part of the dissipative regime where the excitation spectrum
remains discrete (sharp quasiparticle levels), specifically in
layered superconductors\cite{FS,LO98,KL,GP}. In the clean limit
(i.e., for scattering rates $\hbar/\tau$ much smaller than the
superconducting gap $\Delta$), the motion of quasiparticles in the
vortex core is ballistic: they cross the vortex core many times
before scattering off impurities. Therefore in this limit the
spectral properties are sensitive to the details of disorder
realization. In the superclean regime ($\hbar/\tau \ll \Delta^2
/E_F$, where $E_F$ is the Fermi energy) the levels inside a
two-dimensional vortex split into two sets (`combs') of equally
spaced levels \cite{LO98}. The transformation of this correlated
spectrum into a
featureless uncorrelated one with increasing disorder proceeds
in distinct steps: within the moderately clean regime
($\Delta^2/E_F \ll \hbar/\tau \ll \Delta$) a new intermediate
region ($\Delta^2/E_F \ll \hbar/\tau \ll \Delta
\sqrt{\Delta/E_F}$) has been found \cite{LO98,KL} where the
comb structure remains preserved but is randomly shifted in energy
(the number of impurities in the core has to be small enough to
preserve the `combs', while being large enough to randomize the
overall shift). Increasing the impurity concentration further
(already in the moderately clean limit for weak impurities
\cite{SKF}, and ultimately in the dirty limit \cite{Zirnbauer}),
the `comb' structure is destroyed with a crossover to the class
$C$ random-matrix ensemble\cite{AZ}.

Keeping within the moderately clean regime with a discrete
spectrum, in two-dimensional (or layered) superconductors there
further exist two limits of dissipation \cite{FS}: The first one
applies to the slowly moving vortex with a discrete quasiparticle
spectrum (``discrete-spectrum regime'') where the dissipation is
due to Landau--Zener transitions between individual levels
\cite{LZ,Wilkinson}. The second limit is with levels smeared into
a continuum either by the vortex motion or by inelastic processes
(``continuum-spectrum regime''); the dissipation then is given by
the linear-response Kubo formula \cite{Kubo}. At low temperatures,
the inelastic smearing can be neglected and the crossover between
these two regimes is controlled by the vortex velocity with the
characteristic velocity given by $v_K\sim(\Delta/p_F)/\sqrt{k_Fl}$
(here, $\Delta$ is the superconducting gap and $l$ denotes the
elastic mean free path\cite{FS}).

In the framework of random-matrix models with time-dependent
Hamiltonians, the dissipation in the discrete-spectrum and
continuum-spectrum regimes was considered by Wilkinson
\cite{Wilkinson}. He finds that, in the case of the unitary
Wigner--Dyson ensemble, the linear dissipative response remains
valid in the whole range of velocities, both in the
continuum-spectrum (high velocity) and in the discrete-spectrum
(low velocity) regimes. Based on this fact and on the similarity
between the unitary and class $C$ ensembles, it was shown that
with the class $C$ level statistics the dissipation rate nearly
follows the Bardeen-Stephen prediction, even in the limit of small
velocities $v<v_K$ where the quasiclassical description is no
longer valid \cite{FS}. Contrary, it was claimed in
Ref.~\onlinecite{KL} that the additional correlation between
levels (the `two-comb' structure) may lead to an anomalously high
vortex viscosity in the moderately clean regime in the
low-velocity limit.

In our paper, we reconsider the problem of the vortex viscosity in a
two-dimensional $s$-wave superconductor, taking full account of
the discreteness of the vortex spectrum and of the microscopic
structure of the quasiparticle levels. We assume the moderately
clean limit with an appropriate number of impurities in the core,
such that the spectrum possesses the randomly shifted two-comb
structure (the precise condition is given in
Section~\ref{circular-ensemble}). We find that in spite of the
level correlations derived in Refs.~\onlinecite{LO98,KL}, the
vortex viscosity does not differ from the well-known
quasiclassical result
\cite{GK73,Bardeen-Sherman,LO76,Kopnin-Kravtsov,Kopnin94,Kopnin-Lopatin}
for three-dimensional superconductors. Within the same microscopic
model we consider separately the discrete-spectrum and
continuum-spectrum regimes of dissipation: in both limits we
arrive at the Bardeen--Stephen result for the dissipation,
\be
  \sigma_{xx} = \frac{ecn\omega_0\tau}{B} \sim \sigma_n
  \frac{H_{c2}}{B},
\label{sigma-xx}
\ee
where $n$ is the electron density, $\hbar\omega_0\sim\Delta^2/E_F$ is
the spacing between the levels in the core\cite{CDGM}, and $B$ is
the magnetic field. The two-comb level structure found in
Refs.~\onlinecite{LO98,KL} may be classified as the circular
unitary random-matrix ensemble of dimension two\cite{Mehta}.
Therefore our findings may be considered as a generalization of
Wilkinson's results \cite{Wilkinson} to circular ensembles. We
also find the scattering time $\tau$ for quasiparticles in the
core in the $s$-wave approximation: for a weak impurity strength,
the effective $1/\tau$ is larger than the bulk scattering rate
$1/\tau_n$ by the logarithm of the impurity strength. A similar
logarithmic correction was previously derived in
Refs.~\onlinecite{Kopnin94,Kopnin-Lopatin}.

The paper is organized as follows. In Section
\ref{section-hamiltonian} we prepare for the calculation by
deriving the microscopic Hamiltonian projected onto the relevant
subgap states in the vortex core. In Section
\ref{circular-ensemble} we review the results of
Refs.~\onlinecite{LO98,KL} on the circular random-matrix ensemble
appearing in a disordered vortex at the intermediate level
disorder. We then describe the two limiting regimes of dissipation
(Sect. \ref{two-regimes}), the discrete-spectrum and the
continuum-spectrum regimes, and set up the stage for the
calculations. In Section \ref{section-discrete} we treat the case
of the discrete energy spectrum (low velocities and no inelastic
level broadening), while Section \ref{section-continuum} is
devoted to the opposite continuum-spectrum limit where the levels
are broader than the interlevel distance. Finally, we discuss our
findings in Section \ref{section-discussion}. The sensitivity of
the energy levels to the vortex displacement is calculated in an
Appendix.

\section{Microscopic Hamiltonian of the moving vortex}
\label{section-hamiltonian}

Before discussing the moving vortex, let us review the excitation
spectrum of a clean two-dimensional vortex at rest. The
corresponding wave functions are given by the solutions to the
Bogoliubov--deGennes equations
\be
  \begin{pmatrix}
    H_0 & \Delta(\br) \\
    \Delta^*(\br) & -H_0
  \end{pmatrix}
  \begin{pmatrix}u \\ v\end{pmatrix}
  = E
  \begin{pmatrix}u \\ v\end{pmatrix},
\label{H0}
\ee
where $H_0={\bf p}^2/2m-E_F$. We assume an axially symmetric
vortex with the order parameter $\Delta(\br)=\Delta(r)
e^{i\theta}$, where the modulus of the order parameter depends
only on the radial component $r$ and the phase winds with the
angular coordinate $\theta$. We neglect the magnetic field in the
vortex core, assuming a large magnetic penetration depth $\lambda
\gg \xi$, where $\xi=v_F/\pi\Delta$ is the superconducting
coherence length (here and below we choose units with $\hbar=1$).
In the quasiclassical limit $k_F\xi\gg 1$, the spectrum and the
eigenfunctions may be easily found\cite{CDGM}. The eigenvalues
form a spectrum of equidistant levels
\be
  E_\mu=\mu \omega_0,
\ee
where the angular momentum $\mu$ takes half-integer values,
$\mu=n+1/2$, and
\be
  \omega_0 =
  \frac{\int_0^\infty\frac{\Delta(r)}{k_Fr} e^{-2K(r)} \, dr}
     {\int_0^\infty e^{-2K(r)} \, dr},
\ee
\be
  K(r) = \int_0^r \frac{\Delta(r')}{v_F} \, dr';
\ee
$r$ is the distance from the core center. The basic electronic
energy scale in the vortex core takes the value $\omega_0\sim
\Delta^2/\varepsilon_F$, up to a possible logarithmic prefactor
$\log(\Delta/T)$ due to the shrinkage of the vortex core at small
temperatures $T$ (Kramer--Pesch effect\cite{Kramer-Pesch}). The
eigenstates $\Psi_\mu=(u_\mu, v_\mu)^T$ take the form
\begin{equation}
  \Psi_\mu(\br) = {\cal A}
  \begin{pmatrix}
    J_{\mu-1/2}(k_Fr) \, e^{-i(\mu-1/2)\theta} \\
    J_{\mu+1/2}(k_Fr) \, e^{-i(\mu+1/2)\theta}
  \end{pmatrix} e^{-K(r)},
\label{subgap-states-mu}
\end{equation}
where
${\cal A}^2= [4k_F^{-1} \int_0^\infty e^{-2K(r)} dr]^{-1} \sim k_F/\xi$
is the normalization factor, and $J_n(x)$ are the Bessel functions.

In the following, we will be interested in processes at energies
far below the superconducting gap and thus project all the
operators onto the subgap states (\ref{subgap-states-mu}). It will
be convenient to take the Fourier transform of these eigenvectors
in the variable $\mu$ and introduce a new angular variable $\phi$
labelling the direction of the quasiclassical motion of the
quasiparticle,
\be
  \Psi_\phi(\br) \equiv
  \sum_\mu \Psi_\mu(\br) \, e^{i\mu\phi}
  = {\cal A}
  \begin{pmatrix} e^{i\phi/2} \\ e^{-i\phi/2} \end{pmatrix}
  e^{-K(r)} e^{i \bk_\phi\br},
\label{subgap-states-phi}
\ee
where we introduced the vector $\bk_\phi=k_F(\sin\phi,-\cos\phi)$
pointing perpendicular to the direction specified by the angle
$\phi$, with absolute value $k_F$. The plane-wave exponent in
(\ref{subgap-states-phi}) is then $\bk_\phi \br=k_F r
\sin(\phi-\theta)$. This basis of wave functions has a very simple
structure: in addition to the phase winding $\exp(\pm i\phi/2)$
(providing the antiperiodic boundary conditions in $\phi$), these
wave functions are plane waves in the direction of
the wave vector $\bk_\phi$ restricted to a region of size $\xi$
around the vortex center. Note that $\Psi_\phi(\br)$ is not an
eigenfunction of the Hamiltonian (\ref{H0}). Prepared in such a
state at $t=0$, the wave function will rotate in the $\phi$-basis
according to $\Psi(\br,t)=\Psi_{\phi-\omega_0t}(\br)$.

We now turn to the problem of the moving vortex with impurities.
It can be described by the time-dependent Bogoliubov--deGennes
equations
\be
  i\frac{\partial}{\partial t} \Psi=H(t) \Psi,
\ee
where $H(t)$ is the Hamiltonian of the vortex at the position
$\br=\bv t$ with $\bv$ the vortex velocity,
\be
  H(t) = \begin{pmatrix}
    H_0+U(\br) & \Delta(\br-\bv t) \\
    \Delta^*(\br-\bv t) & -H_0-U(\br)
  \end{pmatrix},
\label{moving-vortex-Hamiltonian}
\ee
and $U(\br)$ is the potential set up by the impurities.

In the clean limit ($\Delta\tau\gg1$), the admixture of bulk
states (with energies greater than $\Delta$) to the vortex states
may be neglected. We project the Hamiltonian
(\ref{moving-vortex-Hamiltonian}) onto the subgap states of a
clean vortex (\ref{subgap-states-phi}) by substituting
\be
  \Psi(\br,t) =
  \int_0^{2\pi} \frac{d\phi}{2\pi} a(\phi,t) \Psi_\phi(\br-\bv t).
\ee
Defined in this way, the amplitude $a(\phi,t)$ has antiperiodic
boundary conditions in $\phi$, $a(\phi+2\pi)=-a(\phi)$. The time
evolution of the coefficients $a(\phi,t)$ obeys the
Schr\"odinger-type equation
\be
  i\frac{\partial}{\partial t} a =
  -i\omega_0\frac{\partial}{\partial\phi} a +
  \int \frac{d\phi'}{2\pi} T_{\phi \phi'} a(\phi') +
  \int \frac{d\phi'}{2\pi} L_{\phi \phi'} a(\phi'),
\label{equation-of-motion}
\ee
where the kernel
\begin{align}
T_{\phi \phi'} &=\left\langle \Psi_\phi(\br-\bv t) \right|
  \left( -i\frac{\partial}{\partial t} \right)
  \left| \Psi_{\phi'}(\br-\bv t) \right\rangle  \nonumber\\
&=
  i\bv \left\langle \Psi_\phi\right|
  \nabla
  \left| \Psi_{\phi'}\right\rangle
\label{T-kernel}
\end{align}
is produced by the vortex motion, and the kernel
\be
  L_{\phi \phi'} =\left\langle \Psi_\phi(\br-\bv t) \right|
  U(\br) \tau_z
  \left| \Psi_{\phi'}(\br-\bv t) \right\rangle
\ee
is due to the impurities.

The $T$-kernel may be easily computed from the explicit form
(\ref{subgap-states-phi}) of $\Psi_\phi$. In the limit $k_F\xi\gg 1$,
the matrix element (\ref{T-kernel}) takes the form
\be
  T_{\phi \phi'} = - 2\pi\delta(\phi-\phi') \, \bk_\phi \bv
\ee
and we identify this term with the ``Doppler shift''.

The impurity potential $U(\br)$ is taken as a sum over point-like
impurities,
\be
  U(\br)= \sum_i V_i \,  \delta(\br-\br_i).
\ee
Then the scattering kernel may be expressed as \cite{KL,SKF}
\be
   L_{\phi\phi'}
   \! = \!
   2i {\cal A}^2 \! \sin\frac{\phi' {-} \phi}{2} \,
   \sum_i V_i e^{-2 K(|\br_i-\bv t|)}
   e^{i(\bk_{\phi'}-\bk_{\phi})(\br_i-\bv t)}.
   \label{scattering-L}
\ee
Summarizing, we end up with the equation of motion
(\ref{equation-of-motion}) containing three terms: The first term
describes the circular (``chiral'') motion of the quasiparticle in
the vortex\cite{CDGM,LO98,KL}. The second term is the Doppler
shift due to the vortex motion. And the third term describes the
scattering off impurities (taking point-like impurities is
equivalent to including only $s$-wave scattering). We are
interested in the energy pumping in this time-dependent model.
Note that the physical energy is given by the first and third
terms in the evolution operator (\ref{equation-of-motion}) but
does not include the second term ($T$-term) which arises from the
time-derivative of the basis wave functions. This discrepancy
between the energy and the evolution operator may be resolved by
an appropriate {\it time-dependent gauge transformation}
\be
  \tilde{a}(\phi) = a(\phi) e^{-i\bk_\phi \bv t}.
\label{tilded-basis}
\ee
This gauge transformation has a dual effect on the equation of
motion (\ref{equation-of-motion}): Firstly, it makes the energy
operator coincide with the evolution operator. The old $T$-kernel
is now replaced by a time-dependent one,
\be \tilde{T}_{\phi\phi'} =
  2\pi\delta(\phi-\phi') \:
  \omega_0( \bn_z \times \bk_\phi) \bv t ,
\label{tilded-T}
\ee
where $\bn_z$ is the unit vector perpendicular to the plane (this
gauge transformation resembles the one in electrodynamics
replacing a static electric field with a magnetic field linearly
growing in time). Secondly, the $L$-kernel is transformed as well
and the new kernel takes the form
\be
   \tilde{L}_{\phi\phi'}=
   2i {\cal A}^2 \sin\frac{\phi'-\phi}{2} \, \sum_i V_i e^{-2
   K(|\br_i-\bv t|)} e^{i(\bk_{\phi'}-\bk_{\phi})\br_i} ,
   \label{tilded-L}
\ee
which differs from Eq.~(\ref{scattering-L}) by the cancellation of
the velocity term ${\bf v}t$ in the last exponent.

The physical content of the two gauges may be understood in the
following way: The basis wave functions (\ref{subgap-states-phi})
are quasiclassical plane waves (at the wave vector $k_F$) cut off
by the long-wavelength envelope $\exp(-2K(r))$ of the size of
order $\xi$. In the original gauge [with variables $a(\phi,t)$],
this basis was chosen by simply translating the
basis~(\ref{subgap-states-phi}) together with the vortex. In the
new gauge [with variables $\tilde{a}(\phi,t)$], only the
long-wavelength envelope is translated, without shifting the
phases of the quasiclassical plane waves.

This difference in matching phases of plane waves at different
vortex positions produces two different descriptions of the moving
vortex. In the first gauge [variables $a(\phi,t)$], the evolution
equation (\ref{equation-of-motion}) contains impurities moving
with respect to the vortex (fast oscillations in $L_{\phi\phi'}$).
This description is similar to the approach taken in
Refs.~\onlinecite{FS,LO98,KL}, a static vortex subject to moving
impurities. However, in those references, the $T$-term was
omitted. We will show below that omitting this term does not
change the result for the dissipation in the moderately clean
limit, thus justifying the approach of
Refs.~\onlinecite{FS,LO98,KL}. In the second gauge [variables
$\tilde{a}(\phi,t)$], there are no oscillating terms in
$L_{\phi\phi'}$ (except for the slowly varying envelope $K(r)$
whose time derivative may be neglected in most cases). All the
oscillations in the $L$-term in the first gauge may be removed by
a single gauge transformation, as all impurities move with respect
to the vortex with the same velocity and in the same direction.
This fact has not been properly taken into account in
Ref.~\onlinecite{KL}, which has lead to an unphysical result. We
shall comment in more detail on the derivation of
Ref.~\onlinecite{KL} in Sec.~\ref{section-discrete}; here we just
remark that the parallel motion of the impurities with respect to
the vortex requires special care in the calculations within the
first gauge, but is automatically taken into account in the second gauge.

\section{Circular unitary ensemble of the quasiparticle
levels in the disordered vortex}
\label{circular-ensemble}

In this section, we review the derivation of the
two-comb spectral statistics in the vortex from Refs.~\onlinecite{LO98,KL}
and set up the notation
for the calculations in the later sections. For the discussion of
the spectral statistics in this section we take the vortex at rest
($\bv=0$). Then the Hamiltonian in the equation of motion
(\ref{equation-of-motion}) contains only two terms, the kinetic
term $-i\omega_0\partial/\partial\phi$ and the scattering term
$L_{\phi\phi'}$ (for a vortex at rest there is no difference
between $L_{\phi\phi'}$ and  $\tilde{L}_{\phi\phi'}$).

At not very high impurity concentration [see
Eq.~(\ref{upper-bound}) below], the scattering kernel
$L_{\phi\phi'}$ may be approximated as a sum of two local terms
[this approximation is due to the rapidly oscillating exponent in
(\ref{scattering-L})],
\be
  L_{\phi\phi'} = 4\pi i\omega_0 \sum_{i=1}^N
  [ J_i \check\delta_{\phi,\phi_i} \check\delta_{\phi',\phi_i+\pi}
  - J_i^* \check\delta_{\phi,\phi_i+\pi} \check\delta_{\phi',\phi_i} ] ,
\label{imp-potential-KL}
\ee
where $N$ is the number of impurities in the core, the parameters
$\phi_i$ specify the angular positions of the impurities, and
$\check\delta_{\phi_1,\phi_2}$ are $\delta$-functions smeared over
the width $(\phi_1-\phi_2)\sim (k_F\xi)^{-1/2}$ and
antiperiodically continued in $\phi_1$ and $\phi_2$
($\check\delta_{\phi_1,\phi_2+2\pi}=\check\delta_{\phi_1+2\pi,\phi_2}
=-\check\delta_{\phi_1,\phi_2}$). Note that the regularization
(\ref{imp-potential-KL}) as a product of two smeared
$\delta$-functions is important for evaluating the scattering
matrix (\ref{M-i}) below \cite{LO98,KL}. The effective strength of
the $i$-th impurity is
\be
  J_i =
  \frac{i{\cal A}^2V_i e^{-2K(r_i)+2ik_Fr_i}}{\omega_0 k_F r_i}
  \sim i\vartheta_i \frac\xi{r_i} e^{-2K(r_i)+2ik_Fr_i} ,
\label{J-i}
\ee
where $\vartheta_i=mV_i$ is the Born parameter of the impurity (an
additional imaginary unit compared to the notation in
Ref.~\onlinecite{KL} is due to the antiperiodic boundary
conditions employed).

\begin{figure}
\epsfxsize=0.6\hsize
\centerline{\epsfbox{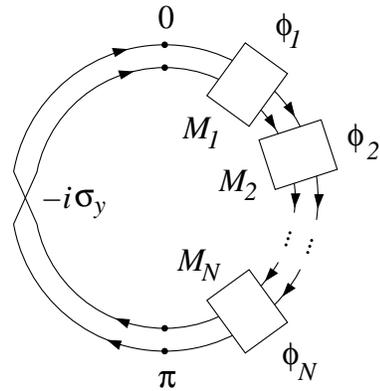}}
\medskip
\caption{Schematic representation of the evolution of the wave
function $\psi(\phi)$ in the circular-unitary-ensemble regime. The
two-component wave function $\psi(\phi)$ rotates with the angular
velocity $\omega_0$ and scatters off impurities. The impurity
scattering is local and is described by the unitary matrices
$M_i$, according to (\ref{bc-imp}). After rotating by half a turn,
the wave function is projected back to the origin with the help of
the boundary condition (\ref{bc-pi}).} \label{fig}
\end{figure}

The matrix elements (\ref{imp-potential-KL}) couple only angles
with difference close to $\pi$. Introducing the two-component
vector
\be
  \psi(\phi) = \begin{pmatrix} a(\phi) \\ a(\phi+\pi) \end{pmatrix},
  \qquad \phi\in [0;\pi],
\label{psi-doublet}
\ee
the scattering becomes local in $\phi$. The individual scattering
events in (\ref{equation-of-motion}) then may be integrated
separately and formulated in terms of a boundary condition
\cite{KL}
\be
  \psi(\phi_i+0)=M_i \psi(\phi_i-0),
\label{bc-imp}
\ee
where
\be
  M_i = \frac{1}{1+|J_i|^2}
  \begin{pmatrix}
    1-|J_i|^2 & 2J_i \\
    -2J_i^* & 1-|J_i|^2
  \end{pmatrix}.
\label{M-i}
\ee
The boundary condition for $\psi(\phi)$ going around the half
circle is
\be
  \psi(0)=-i\sigma_y \psi(\pi).
\label{bc-pi}
\ee
Thus the energy levels are determined by the eigenvalues of the
matrix $-i\sigma_y M$, where
\be
  M = M_N \dots M_2 M_1,\quad
  \pi>\phi_N > \dots > \phi_1 > 0.
\label{M-def}
\ee
The full ``scattering matrix'' $-i\sigma_y M$ is a unitary matrix
with the eigenvalues $\exp(\pm i\pi E/\omega_0)$, so that the
energy levels $E$ are solutions to the equation
\be
  \cos(\pi E/\omega_0)
  = -\frac i2 \tr \sigma_y M.
\label{KL-spectrum}
\ee
The evolution of the wave function $\psi(\phi)$ is
schematically shown in Fig.~\ref{fig}.

In the moderately clean limit with a sufficiently large (but not
too large) number of impurities in the core, the matrix $M$ (and
thus also $-i\sigma_y M$) is random with a uniform distribution
over the $SU(2)$ group. Such a random-matrix ensemble is
classified as the ``circular unitary'' ensemble with dimension two
\cite{Mehta}. The density of states and the level correlations may
be easily computed from Eq.~(\ref{KL-spectrum}). For example, the
average density of states is given by\cite{KL}
\be
  \corr{\rho(E)} =
  \frac2{\omega_0} \sin^2 \left( \pi\frac{E}{\omega_0} \right) .
\label{KL-DOS}
\ee
The spectrum consists of two combs of equidistant levels: A level
is characterized by the comb number $s=\pm1$ and its position in
the comb, $E_{\pm,k} = \pm (E_0 + 2\omega_0 k)$, where $E_0$ is
the eigenvalue of the lowest level with positive energy ($0\leq
E_0<\omega_0$). The eigenfunctions can be written explicitly as
\be
  \psi_{\pm,k}(\phi) =
  e^{i\phi E_{\pm,k} /\omega_0} M_{\phi,0} \psi_{\pm}(\phi=0),
\label{Psi-f}
\ee
where $\psi_{\pm}(\phi=0)$ are the two eigenvectors of $-i\sigma_y
M$, and the scattering matrices $M_{\phi,\phi'}$ are defined as
\be
  M_{\phi,\phi'} =
  M_{i_n} \dots M_{i_1},
  \quad
  \phi>\phi_{i_n} > \dots > \phi_{i_1} > \phi',
  \label{M-two-point}
\ee
where the product is taken over all impurities between the angles
$\phi'$ and $\phi$. Eq.~(\ref{M-two-point}) introduces
$M_{\phi,\phi'}$ for $\phi>\phi'$; furthermore, it is convenient
to define the scattering matrix for $\phi<\phi'$ as
$M_{\phi,\phi'}=M_{\phi',\phi}^\dagger$. The symmetry of the
$SU(2)$ matrices
\be
  M_{\phi,\phi'}^* = \sigma_y M_{\phi,\phi'} \sigma_y
\label{M-symmetry}
\ee
allows us to translate easily between the eigenfunctions
for the two series of levels via
\be
  \psi_{-,-k}(\phi) = \sigma_y \psi_{+,k}^*(\phi) .
\label{psi-symmetry}
\ee

In our calculations, we shall need not only the properties of the
spectrum, but also some statistical properties of the wave
functions in the regime of the circular unitary ensemble. At any
given point $\phi$ and for any energy level, the wave function
$\psi(\phi)$ is a spinor pointing in a random direction. All
directions are equally probable, allowing us to compute
equal-point correlation functions such as
\begin{subequations}
\label{ff}
\begin{align}
  \langle \psi^\dagger(\phi) \sigma_\alpha \psi(\phi) \cdot
  \psi^\dagger(\phi) \sigma_\beta \psi(\phi) \rangle &=
  \frac{\delta_{\alpha\beta}}{3},
\label{ff1}
\\
  \langle \psi^\dagger(\phi) \sigma_\alpha \psi^*(\phi) \cdot
  \psi^T(\phi) \sigma_\beta \psi(\phi) \rangle &=
  \frac{2 \delta_{\alpha\beta}}{3}.
\label{ff2}
\end{align}
\end{subequations}

The correlations between wave functions at different values of
$\phi$ may be expressed in terms of the properties of the
scattering matrices (\ref{M-two-point}). The angular correlations
between scattering matrices decay exponentially,
\be
  \corr{
  M^\dagger_{\phi,\phi'} \sigma_z M_{\phi,\phi'}}
  =
  e^{-|\phi-\phi'|/\phi_\tau} \sigma_z,
  \label{wf-decay}
\ee
where
\be
  \frac{1}{\omega_0\tau} \equiv
  \frac{1}{\phi_\tau}
  = 16n_{\rm imp} \int_0^\infty r\,dr \frac{|J(r)|^2}{(1+|J(r)|^2)^2},
\label{phitau}
\ee
and $n_{\rm imp}$ is the impurity concentration. In (\ref{phitau})
we have assumed (for simplicity) that all impurities have equal
Born parameters, in which case the effective impurity strength
$J_i$ as given by (\ref{J-i}) becomes a function of $r_i$ and is
denoted $J(r)$ in (\ref{phitau}). In order to derive
(\ref{wf-decay}) and (\ref{phitau}) one takes a small increment of
$\corr{M^\dagger_{\phi,\phi'} \sigma_z M_{\phi,\phi'}}$ in $\phi$
and averages over a single impurity scattering matrix,
\begin{multline}
  \frac{\partial}{\partial\phi}
  \corr{M^\dagger_{\phi,\phi'} \sigma_z M_{\phi,\phi'}}
  = 2 n_{\rm imp} \int_0^\infty r_i\, dr_i \\
  {} \times
  \left[
    \corr{M^\dagger_{\phi,\phi'} M_i^\dagger \sigma_z M_i M_{\phi,\phi'}}
    - \corr{M^\dagger_{\phi,\phi'} \sigma_z M_{\phi,\phi'}}
  \right] .
\end{multline}
The last averaging is performed independently over
$M_{\phi,\phi'}$ and $M_i$. In averaging, we first account for the
rapidly oscillating off-diagonal elements in (\ref{M-i}) and
arrive at (\ref{wf-decay}), (\ref{phitau}).

For strong impurities with the Born parameter $\vartheta\equiv
mV_i\sim1$ (as above, we assume that all impurities are of equal
strength), the scattering time $\tau$ is of the order of the bulk
normal-state elastic mean-free time $\tau_n$. For weak impurities
with $\vartheta\ll 1$, the integral in Eq.~(\ref{phitau}) has a
logarithmic dependence on the Born parameter\cite{KL},
\be
  \frac 1\tau \sim \frac 1{\tau_n} \ln \frac1\vartheta.
\label{1/tau}
\ee
A similar logarithmic behavior of the scattering rate was also
derived in Refs.~\onlinecite{Kopnin94,Kopnin-Lopatin}. The
relaxation angle $\phi_\tau$ in (\ref{phitau}) plays a central
role in the context of vortex dissipation, as it enters the
expression for the friction coefficient and hence also for the
flux flow conductivity. In Ref.~\onlinecite{KL} this quantity
entered the {\em numerator} in the expression for the
conductivity, while in our results it enters the {\em denominator}
in the Bardeen--Stephen form (\ref{sigma-xx}). We comment in
detail on our disagreement with Ref.~\onlinecite{KL} in
Section~\ref{section-discrete} and in the Appendix.

The physical meaning of $\phi_\tau$ is the correlation length of
the scattering matrix $M_{\phi,\phi'}$ and $\tau$ is the
corresponding scattering time. We shall see later that it is the
correlation function (\ref{wf-decay}) which determines the rate of
interlevel transitions. In principle, one can derive a similar
exponential decay ``in the $x$ and $y$ directions'' [with
$\sigma_z$ replaced by $\sigma_x$ or $\sigma_y$ in
(\ref{wf-decay})], in which case the correlation length
$\phi_\tau^{(xy)}$ is twice larger than $\phi_\tau$ ``in the $z$
direction''. This is due to the specific form (\ref{M-i}) of the
scattering matrix $M_i$.

In order to realize the regime of the circular unitary
random-matrix ensemble it is necessary that $\phi_\tau\ll \pi$,
which defines the lower bound for the moderately clean regime,
$\omega_0\tau \ll 1$. However, the region of the circular unitary
ensemble does not extend over the whole moderately clean regime
\cite{KL}. The additional restriction originates from the breaking
of the instant-scattering approximation (\ref{imp-potential-KL})
in the limit of too strong disorder. Indeed, the typical width of
the smeared $\check\delta$-functions in this equation can be
estimated as $\delta\phi \sim 1/\sqrt{k_F\xi}$ (that corresponds
to an impurity at the distance of the order of $\xi$ from the
vortex center). On the other hand, the number of impurities in the
core is $N\sim n_{\rm imp}\xi^2 \sim 1/\omega_0\tau_n\vartheta^2$.
The solution for the spectrum discussed above is justified as long
as the $\check\delta$-functions in Eq.~(\ref{imp-potential-KL}) do
not overlap, which is equivalent to the condition $N\,\delta\phi
\ll 1$. Thus, the circular unitary ensemble is realized only in a
relatively narrow range of disorder strengths
\be
  \frac{\omega_0}{\ln 1/\vartheta} \ll \frac1{\tau_n} \ll
  \vartheta^2 \sqrt{\Delta\omega_0}
\label{upper-bound}
\ee
[we assume that the Born parameter $\vartheta\ll 1$ and make use
of Eq.~(\ref{1/tau}); for $\vartheta\sim 1$ one should replace
$\ln 1/\vartheta$ by 1]. For stronger disorder (i.e., larger
scattering rate $1/\tau_n$) or smaller Born parameter $\vartheta$ the
instant-scattering approximation (\ref{imp-potential-KL}) fails
and the circular unitary ensemble crosses over to the class $C$
ensemble \cite{Zirnbauer,SKF}. A phenomenological approach to this
crossover has been discussed in Ref.~\onlinecite{BHL} and the
results of numerical simulations are available in
Ref.~\onlinecite{Fujita}.

\section{Two regimes of dissipation}
\label{two-regimes}

The time evolution of the core states is described by
Eq.~(\ref{equation-of-motion}). Since the evolution operator is
nonstationary (i.e., explicitly time-dependent), it causes
transitions between levels with different energy. In a fermionic
system, the rate of downward transitions is suppressed compared to
the rate of upward transitions due to the Pauli exclusion
principle, leading to an increase of the average energy with time.
The energy pumped into the system will finally be transferred to
the thermal bath via the interaction with phonons or other soft
degrees of freedom, thus producing a finite dissipation.

There are two different mechanisms of dissipation depending on
whether the individual energy levels can be resolved or not
\cite{Wilkinson}. If the discrete spectrum is smeared into a
continuous one, the energy pumping can be calculated with the help
of the standard linear-response Kubo formula \cite{Kubo}. This
regime is naturally realized when the {\em inelastic width}
$\gamma_{\rm in}$ of an energy level exceeds the mean level
spacing $\omega_0$. On the other hand, the spectrum may turn out
effectively continuous even at $\gamma_{\rm in}\ll\omega_0$ if the
time dependence of the evolution operator on the right-hand side
of Eq.~(\ref{equation-of-motion}) is so fast that it destroys the
instantaneous adiabatic spectrum \cite{Wilkinson}. In this case,
the frequency $\gamma_v$ of perturbations due to the
nonadiabaticity of the spectrum exceeds $\omega_0$ and plays the
role of the effective level width. In the opposite case, when
$\gamma_{\rm in}, \gamma_{v} \ll \omega_0$, the spectrum is
essentially discrete and the dissipation is due to rare
Landau--Zener transitions \cite{LZ} taking place when two levels
come very close to each other.

In a normal metal at low temperatures, the inelastic widths due to
the electron-electron and electron-phonon interactions are given
by $\gamma_{e-e}\sim T^2/E_F$ and $\gamma_{e-ph}\sim
T^3/\Theta_D^2$, respectively\cite{inelastic} ($\Theta_D$ is the
Debye temperature). Since $\gamma_{e-e}$ is of the order of
$\omega_0$ already at $T\sim\Delta$, it does not contribute to the
level smearing at lower temperatures. Furthermore, for the states
localized in the vortex core the interaction with phonons appears
to be strongly suppressed compared to the normal-state rate
\cite{e-ph}; one of the main reasons is that the quasiparticles
with energy $\mu\omega_0\ll\Delta$ are composed of nearly equal
mixtures of electron and hole components [i.e., $\corr{u_\mu^2}
\approx \corr{v_\mu^2}$, cf.\ Eq.~(\ref{subgap-states-mu})] with
negligible net charge. Thus, at sufficiently small temperatures
$T\ll\Delta$, the inelastic width of the core states $\gamma_{\rm
in}\ll\omega_0$ and the regime of dissipation is determined solely
by the vortex velocity. In Ref.~\onlinecite{FS} the crossover
velocity $v_K$ separating the regimes with discrete and continuous
spectra was estimated as
\be
  v_K \sim \frac{\Delta/p_F}{\sqrt{k_Fl}}
\label{v_K}
\ee
and we will present a microscopic derivation of this result below.

For model Hamiltonians from the three Wigner-Dyson random-matrix
ensembles \cite{Mehta}, dissipation was calculated by Wilkinson
\cite{Wilkinson,Wilkinson-comment} in the regimes of small ($v\ll
v_K$) and large ($v\gg v_K$) velocities (in the limit $\gamma_{\rm
in}=0$). He finds that if the Hamiltonian $H(X)$ of the system
depends on time through $X=vt$ then the energy dissipation is
determined by the variance of the (off-diagonal, $i\neq k$) matrix
elements $C(0) \equiv \Corr{|\partial H_{ik}/\partial
X|^2}/\omega_0^2$ normalized by the mean level spacing $\omega_0$.
In the continuous-spectrum regime specified by $v\gg v_K
\equiv\omega_0C^{-1/2}(0)$, the energy dissipation rate as given
by the Kubo formula,
\be
  W_{\rm Kubo}=\pi C(0)v^2,
\label{W-Kubo}
\ee
is the same for the orthogonal ($\beta=1$), unitary ($\beta=2$),
and symplectic ($\beta=4$) ensembles and describes viscous
damping. In the discrete-spectrum regime, at $v\ll v_K$, the
result depends on the level-repulsion parameter $\beta$, which
determines the probability $P(\varepsilon)\propto
\varepsilon^{\beta}$ to find two levels at a distance $\varepsilon
\ll\omega_0$. In this situation the energy dissipation rate is
given by the expression \cite{Wilkinson}
\be
  W_\beta\propto v^{(\beta+2)/2},
\label{W-Zener}
\ee
hence the dissipation is superohmic for the Gaussian orthogonal
ensemble, while for the Gaussian unitary ensemble it remains ohmic
with $W_2=\pi C(0)v^2$, exactly coinciding with $W_{\rm Kubo}$
despite a very different mechanism of dissipation.

In the next
sections we calculate the dissipation rate for the vortex motion
in the regime of the Koulakov--Larkin `two-comb' spectrum~\cite{KL},
both in the limits of small and large velocities,
thus extending Wilkinson's considerations to the case of circular
unitary ensembles.

\section{Dissipation in the discrete spectrum (Landau--Zener regime)}
\label{section-discrete}

In this section we calculate the dissipation for the two-comb
level structure described in Section~\ref{circular-ensemble} at
small velocities. We will work in the `tilded' basis $\tilde
a(\phi,t)$ introduced in Eq.~(\ref{tilded-basis}), where the
energy coincides with the expectation value of the evolution
operator (\ref{equation-of-motion}), with $T_{\phi\phi'}$ and
$L_{\phi\phi'}$ substituted by their tilded counterparts
(\ref{tilded-T}) and (\ref{tilded-L}). In the low-velocity limit,
the energy is pumped into the system when two levels come very
close to each other and a nonadiabatic Landau--Zener transition
becomes possible. Due to the symmetry of the two-comb spectrum the
rate $R_{\rm\scriptscriptstyle LZ}$ of such transitions is the
same for each neighboring pair
of levels. We will calculate $R_{\rm\scriptscriptstyle LZ}$
by considering the lowest level
with positive energy $\varepsilon$ and its mirror image with
energy $-\varepsilon$. For simplicity we assume that the vortex
moves in the $x$ direction, $v_y=0$.

Following the logic of Ref.~\onlinecite{Wilkinson}, we diagonalize
the Hamiltonian at $t=0$ and restrict it to the $2\times2$ matrix
involving the pair of states considered,
\be
  \tilde H(t) = \begin{pmatrix}
    \eps + h_{11} t & h_{12} t \\
    h_{12}^* t & -\eps - h_{11} t
  \end{pmatrix} ,
\ee
where $h_{ij}$ are the elements of the matrix $\partial H/\partial
t$ (neglecting the quadratic term $\propto(\partial^2H/\partial
t^2)t^2$ is justified in the Landau--Zener regime since the
duration of a nonadiabatic transition is proportional to the gap of
the avoided crossing and is small at $v\ll v_K$). The
instantaneous adiabatic spectrum takes the form
\be
  E(t)
  = \pm\sqrt{\Delta_{\rm\scriptscriptstyle LZ}^2+A^2
  (t-t_{\rm\scriptscriptstyle LZ})^2},
\label{EZener}
\ee
where
\begin{align}
  \Delta_{\rm\scriptscriptstyle LZ}^2
  = \frac{\eps^2|h_{12}|^2}{A^2}, \quad
  A^2 = h_{11}^2+|h_{12}|^2, \quad
  t_{\rm\scriptscriptstyle LZ} = - \frac{\eps h_{11}}{A^2}.
\end{align}

Equation (\ref{EZener}) describes an avoided crossing with the
minimal distance (Landau--Zener gap) $2\Delta_{\rm
\scriptscriptstyle LZ}$ between the spectral branches realized at
$t=t_{\rm\scriptscriptstyle LZ}$. The probability of the
Landau--Zener transition at such a crossing is
$\exp(-\pi\Delta_{\rm\scriptscriptstyle LZ}^2/A)$. The mean rate
of transitions is given by
\be
  R_{\rm\scriptscriptstyle LZ} = \Corr{\delta(t_{\rm\scriptscriptstyle LZ})
  \exp\left(-\pi\frac{\Delta_{\rm\scriptscriptstyle LZ}^2}{A}\right)},
\label{R-Zener}
\ee
where the role of $\delta(t_{\rm\scriptscriptstyle LZ})$ is to
count each avoided crossing once. The average in
Eq.~(\ref{R-Zener}) is taken over the distribution of the
parameters $\eps$, $h_{11}$, and $h_{12}$ describing the avoided
crossing. The energy $\eps$ is expressed through the transfer
matrix $M$ according to Eq.~(\ref{KL-spectrum}). The coefficients
$h_{11}$ and $h_{12}$ are the matrix elements of the perturbation
(\ref{tilded-T}) over the exact wave functions $\psi_\pm(\phi)
\equiv \psi_{\pm,0}(\phi)$ [which depend on the trajectory
$M_{\phi,0}$ via Eq.~(\ref{Psi-f})],
\begin{align}
  h_{11} = \omega_0 k_F v \int_0^\pi \frac{d\phi}\pi
    \psi_+^\dagger(\phi) \sigma_z \psi_+(\phi) \cos\phi ,
\\
  h_{12} = \omega_0 k_F v \int_0^\pi \frac{d\phi}\pi
    \psi_+^\dagger(\phi) \sigma_z \psi_-(\phi) \cos\phi .
\end{align}

We come to the crucial point: The quantities $\eps$, $h_{11}$, and
$h_{12}$ have different dependencies on the transfer matrix
$M_{\phi,0}$. In the moderately clean limit, when the number of
impurities in the core is sufficiently large and $\phi_\tau\ll 1$,
the matrix $M_{\phi,0}$ performs many rotations over the $SU(2)$
group. Therefore, we conclude that (i) $\eps$, $h_{11}$, and
$h_{12}$ are uncorrelated, and (ii) the distribution of $h_{11}$
and $h_{12}$ is Gaussian. We further calculate the variances
of $h_{11}$ and $h_{12}$.
For $\corr{h_{11}^2}$ we obtain with the help of
Eq.~(\ref{wf-decay}),
\begin{align}
  \corr{h_{11}^2} = {}& (\omega_0 k_F v)^2
  \int_0^\pi \frac{d\phi}\pi \frac{d\phi'}\pi
  \cos\phi \cos\phi' e^{-|\phi-\phi'|/\phi_\tau}
\nonumber \\
  & {} \times \corr{
    \psi_+^\dagger(\phi) \sigma_z \psi_+(\phi) \cdot
    \psi_+^\dagger(\phi) \sigma_z \psi_+(\phi)} .
\label{var(h11)}
\end{align}
Integrating over $\phi-\phi'$ and averaging over
the spinor $\psi(\phi)$ according to (\ref{ff1}) we find
\be
  \frac{\corr{h_{11}^2}}{(\omega_0 k_F v)^2}
  = \frac{\phi_\tau}\pi
    \Corr{ (\psi_+^\dagger \sigma_z \psi_+)^2 }
  = \frac{\phi_\tau}{3\pi} .
\ee
Analogously, with the help of Eq.~(\ref{ff2}) we obtain
\be
  \frac{\corr{|h_{12}|^2}}{(\omega_0 k_F v)^2}
  = \frac{\phi_\tau}\pi
    \Corr{ |\psi_+^\dagger \sigma_x \psi_+^*|^2 }
  = \frac{2\phi_\tau}{3\pi} .
\ee
In the same way it can be shown that $\corr{h_{11}h_{12}}=0$, thus
proving the statistical independence of $h_{11}$ and $h_{12}$. We
conclude that $h_{11}$, $\Re h_{12}$, and $\Im h_{12}$ are
independently distributed with the same distribution
\be
  {\cal P}(h) = \sqrt{\frac{3}{2\phi_\tau}} \frac1{\omega_0k_Fv}
  \exp\left(-\frac{3\pi h^2}{2\phi_\tau (\omega_0k_Fv)^2}\right) .
\label{P(h)}
\ee
The distribution function for $\eps$ is provided by the density of
states~(\ref{KL-DOS}), in the limit $\eps\ll\omega_0$, ${\cal P}
(\eps) = 2\pi^2 \eps^2/\omega_0^3$.

Averaging Eq.~(\ref{R-Zener}) over $\eps$, $h_{11}$, and $h_{12}$
with the distributions ${\cal P}(\eps)$ and ${\cal P}(h)$, we
obtain for the mean rate of Landau--Zener transitions
\be
  R_{\rm\scriptscriptstyle LZ} = \frac{k_F^2v^2\phi_\tau}{2\omega_0}.
\ee
The energy dissipation rate is given by
$W=\omega_0 R_{\rm\scriptscriptstyle LZ} \equiv \eta
v^2$ with the vortex viscosity
\be
  \eta = \frac12 k_F^2\phi_\tau = \frac12 k_F^2\omega_0\tau .
\label{eta-low-v}
\ee
As defined by Eq.~(\ref{eta-low-v}), $\eta$ is a two-dimensional
viscosity. In a layered superconductor, it determines the friction
force ${\bf F}=-\eta\bv$ exerted on the vortex from excitation of
quasiparticles within one layer.

The statistics (\ref{P(h)}) of the matrix elements of $\partial
H/\partial t$ determines the sensitivity of the spectrum to the
vortex motion and allows one to find the critical velocity $v_K$
separating the regimes of discrete and continuum spectra. The
discrete spectrum can be resolved if the change of the Hamiltonian
during the time $\omega_0^{-1}$, $\delta H \sim h/\omega_0$, is
smaller than $\omega_0$. Taking for $h$ the dispersion of the
distribution (\ref{P(h)}), we find for the critical vortex
velocity
\be
  v_K \sim \frac{\omega_0}{k_F\sqrt{\phi_\tau}},
\label{v_K2}
\ee
coinciding with the estimate (\ref{v_K}) found in
Ref.~\onlinecite{FS}.

Comparing with Ref.~\onlinecite{KL}, we find that our result
(\ref{eta-low-v}) for the viscosity is smaller by a factor
$\phi_\tau^2$ and the expression (\ref{v_K2}) for $v_K$ is larger
by the factor $1/\phi_\tau$. In Ref.~\onlinecite{KL}, following
the analysis in Refs.~\onlinecite{FS,LO98}, it was assumed that
the dissipation is produced by impurities moving with respect to the
vortex, equivalent to neglecting the Doppler-shift term
$T_{\phi\phi'}$ in Eq.~(\ref{equation-of-motion}). In order to
identify the origin of the discrepancy in the results
and verify the validity of
omitting the Doppler shift, we reconsider below the treatment of
Ref.~\onlinecite{KL} --- we will see that the source of the
disagreement is not in the neglect of the Doppler-shift term but
in the neglect of cross-correlations between the motion of different
impurities. To demonstrate this, we adopt the approach
of Ref.~\onlinecite{KL} (omitting the Doppler-shift term) and
re-calculate the correlator of the spatial
gradients of the matrix $M$ \cite{KL-comment1}
\be
  {\cal C}
  = \frac 14
    \biggl<\left(\tr\frac{\partial M}{\partial x}\right)^2 \biggr>,
    \label{eta-KL}
\ee
from which the viscosity follows via $\eta = 2\,{\cal C}/\pi$.
Comparison between Eqs.~(\ref{eta-KL}) and (\ref{W-Kubo}) shows
that the quantity $(2/\pi^2){\cal C}$ plays the role of the
correlation function $C(0)$.

The matrix $M$ specified in Eq.~(\ref{M-def}) is a product of
transfer matrices of individual impurities. Therefore, there are
two contributions to the correlator (\ref{eta-KL}), one
originating from derivatives over the coordinates of the same
impurity (diagonal part) and a second one from derivatives taken
on different impurities (off-diagonal part),
\bea
  {\cal C}_{\rm diag} = \frac 14 \sum_i
    \left< \tr \frac{\partial M}{\partial x_i}
    \tr \frac{\partial M}{\partial x_i} \right> ,
\\
  {\cal C}_{\rm off-diag} = \frac 14 \sum_{i\neq j}
    \left< \tr \frac{\partial M}{\partial x_i}
    \tr \frac{\partial M}{\partial x_j} \right> .
\label{Coffdiag-def}
\eea
The diagonal part is given by \cite{KL}
\be
  {\cal C}_{\rm diag} = \frac{\pi k_F^2}{4\phi_\tau} .
\label{Cdiag}
\ee
The cross term, expressing the correlated nature of the motion of
the impurities with respect to the vortex, has been missed in
Ref.~\onlinecite{KL}. Its calculation is presented in the Appendix
and the result takes the form
\be
  {\cal C}_{\rm off-diag}
  = - \frac{\pi k_F^2}{4\phi_\tau(1+\phi_\tau^2)} .
\label{Coffdiag}
\ee
We thus see that the two contributions ${\cal C}_{\rm diag}$ and
${\cal C}_{\rm off-diag}$ nearly cancel each other and the net
sensitivity of the spectrum to the vortex motion appears to be
significantly lower than that found in Ref.~\onlinecite{KL},
\be
  {\cal C} = \frac{\pi k_F^2 \phi_\tau}{4} ,
\ee
where we employed the condition $\phi_\tau\ll 1$ of the moderately
clean limit. Using Eq.~(\ref{eta-KL}) one recovers the result
(\ref{eta-low-v}) for the vortex viscosity. The above
consideration not only corrects the result of
Ref.~\onlinecite{KL}; in addition, it serves as a microscopic
justification of the model adopted in
Refs.~\onlinecite{FS,LO98,KL}, where the dissipation is due to
impurities moving through the core and the Doppler-shift term $T$
in Eq.~(\ref{equation-of-motion}) is neglected.

\section{Dissipation in the continuum spectrum (Kubo formula)}
\label{section-continuum}

If the vortex velocity $v$ exceeds $v_K$ the discrete spectrum is
smeared and transitions between non-nearest levels become
possible. In this limit, the energy dissipation can be calculated
with the help of the Kubo formula; as in
Sec.~\ref{section-discrete} we will use the `tilded' basis
(\ref{tilded-basis}). Neglecting the slow time dependence of
$\tilde L_{\phi\phi'}$, we write the evolution operator as $\tilde
H(t) = H_0 + It$, where $H_0$ is the Hamiltonian of the vortex at
rest and $I=\tilde T/t$ takes the form
\be
  I_{\phi\phi'} = 2\pi \delta(\phi-\phi') I(\phi), \qquad
  I(\phi) = \omega_0 k_F v \cos\phi ,
\label{I}
\ee
where we again assumed that $v_y=0$. The energy dissipation rate
\be
  W \equiv \frac{dE}{dt} =
  \langle t | \frac{\partial}{\partial t} \tilde H(t) | t \rangle
\ee
is calculated as the linear response to the term $It$ with the
help of the Kubo formula
\be
  W =
  i \int_{-\infty}^0 t' dt' \corr{ 0 |
  [{\cal I}(t'),{\cal I}(0)] | 0} ,
\ee
where
\be
  {\cal I}(t)=\int_0^{2\pi} I(\phi)
  \hat\Psi^\dagger(\phi,t) \hat\Psi(\phi,t) \frac{d\phi}{2\pi}
\ee
and $\hat\Psi(\phi,t)$ is a second-quantized operator in the
interaction representation. The latter can be rewritten in terms
of the (exact disorder dependent) eigenfunctions $\tilde
a_n(\phi)$ and eigenvalues $E_n$ of the Hamiltonian $H_0$,
\be
  \hat\Psi(\phi,t) = \sum_n \hat c_n \tilde a_n(\phi) e^{-iE_nt},
\ee
with $\hat c_n$ the corresponding annihilation operators.
Averaging over the initial state of the vortex at rest (so that
$\tilde a_n(\phi)$ coincides with $a_n(\phi)$) one arrives at the
expression
\bea
  W
  = i \int_{-\infty}^0 t' dt'
    \int \frac{d\phi d\phi'}{(2\pi)^2}
    I(\phi) I(\phi')
    \sum_{km}
    (n_k-n_m)
\nonumber \\
  {} \times
    a_k(\phi) a_k^*(\phi') a_m(\phi') a_m^*(\phi)
    e^{i(E_k-E_m)(t'-t)} ,
\eea
where $n_k=\corr{\hat c_k^\dagger \hat c_k}$ is the distribution
function at the energy $E_k$. This expression can readily be
represented (cf., e.g., Ref.~\onlinecite{Mahan}) in terms of the
Green functions
\be
  G^{R,A}_E(\phi,\phi')
  = \sum_n \frac{a_n(\phi)a_n^*(\phi')}{E-E_n\pm i\delta} ,
\ee
\bea
  W
  = \frac12 \int \frac{d\phi d\phi'}{(2\pi)^2}
    \int \frac{dE}{2\pi} \frac{\partial n(E)}{\partial E}
    I(\phi) I(\phi')
\nonumber \\
  {} \times
    [G^R_E(\phi\phi')-G^A_E(\phi\phi')]
    [G^R_E(\phi'\phi)-G^A_E(\phi'\phi)] .
\label{W-GG}
\eea
The above formula for the dissipation rate is not restricted to
the regime of the circular unitary ensemble but is valid in the
whole moderately clean limit regime as long as $v\gg v_K$. In
particular, one can easily recover the result of the standard
quasiclassical analysis by assuming the $\tau$-approximation:
within this approach the Green function is diagonal in the
momentum representation,
\be
  \corr{G^{R,A}_E(\mu)}
  = \frac{1}{E-\omega_0\mu \pm \frac{i}{2\tau}} ,
\ee
and evaluating the integral in (\ref{W-GG}) one finds the energy
dissipation rate $W=\eta_{\tau} v^2$ with the vortex viscosity
coefficient
\be
  \eta_{\tau} = \frac12 k_F^2 \omega_0\tau .
\label{eta-tau}
\ee

The $\tau$-approximation is the simplest approach to the problem
where all spectral correlations are neglected. Below, we will
calculate the vortex viscosity for the case of the Koulakov-Larkin
two-comb statistics and check how the presence of the two-comb
correlations in the energy levels modifies the result
(\ref{eta-tau}). To this end we rewrite Eq.~(\ref{W-GG}) in terms
of the wave functions $\psi_{sk}(\phi)$ introduced in
(\ref{Psi-f}),
\begin{align}
& W = - \frac12 \sum_{s,s',k,k'}
    \int \frac{d\phi d\phi'}{\pi^2}
    \int \frac{dE}{2\pi} \frac{\partial n(E)}{\partial E}
    \int dt dt'
    e^{i\Phi}
\label{W-Kubo-1} \\
&{}
  \times
    I(\phi) I(\phi')
    \corr{
    \psi_{s'k'}^\dagger(\phi) \sigma_z \psi_{sk}(\phi)
    \cdot
    \psi_{sk}^\dagger(\phi') \sigma_z \psi_{s'k'}(\phi')
    } ,
\nonumber
\end{align}
where $\Phi = E(t-t')-E_{sk}t+E_{s'k'}t'$. We first re-express the
wave functions $\psi_{sk}(\phi)$ via (\ref{Psi-f}) and perform the
summation over $k$ and $k'$ according to the summation formula
\be
   \sum_k e^{i E_{sk}t}
   = \frac{\pi}{\omega_0} e^{i E_{s,0}t}
   \sum_{m=-\infty}^\infty \delta(t-m \pi/\omega_0)
\ee
(which produces a double infinite sum of $\delta$-functions of $t$
and $t'$). We cut off the integrals over $t$ and
$t'$ at some time scale smaller than $\pi/\omega_0$ --- such a cut-off
is equivalent to assuming a smearing of the energy levels
with widths larger than $\omega_0$. Under this assumption,
only one term of the double infinite sum survives with
$t=t'=(\phi-\phi')/\omega_0$. Next, the integrations over $t$,
$t'$, and $E$ can be trivially performed. Finally, we sum over $s$
and $s'$ and arrive at
\begin{multline}
    \corr{
    \psi_{s'}^\dagger(0) \, M^\dagger_{\phi,0} \sigma_z
    M_{\phi,0} \, \psi_{s}(0)
    \psi_{s}^\dagger(0) \, M^\dagger_{\phi',0}\sigma_z
    M_{\phi',0} \, \psi_{s'}(0)} \\
=
\tr\corr{
M^\dagger_{\phi',\phi} \sigma_z M_{\phi',\phi} \sigma_z}.
\end{multline}
This average is calculated with the help of Eq.~(\ref{wf-decay})
yielding
\begin{multline}
  \eta = \frac{k_F^2}{4\pi} \int_0^\pi d\phi d\phi'\,
  \cos\phi\, \cos\phi'\,
  \tr\corr{M^\dagger_{\phi',\phi} \sigma_z M_{\phi',\phi} \sigma_z} \\
  =
  \frac12 k_F^2 \omega_0\tau .
  \label{eta-high-v}
\end{multline}
The form of the above result coincides with (\ref{eta-tau})
calculated within the $\tau$-approximation. The microscopic
expression for the elastic relaxation time $\tau$ is given by
Eq.~(\ref{phitau}). In case the inelastic relaxation time is
shorter than $\tau$, it substitutes the latter in
Eq.~(\ref{eta-high-v}).

To conclude this section we mention that the same result
(\ref{eta-high-v}) can be obtained in the model where the
Doppler-shift term $T_{\phi\phi'}$ in
Eq.~(\ref{equation-of-motion}) is neglected and the dissipation is
due to the motion of impurities with respect to the vortex. Within
that model the energy dissipation rate is calculated as a linear
response to a change in the impurity positions. The derivation
formally repeats the one presented above but with the operator
$I_{\phi\phi'}=v\, \partial L_{\phi\phi'}/\partial x$. Performing
the same manipulations leading to Eqs.~(\ref{W-Kubo-1}) and
(\ref{eta-high-v}) we obtain
\be
  \eta
  = \frac{1}{4\pi}
    \Corr{\tr \frac{\partial M^\dagger}{\partial x}
      \frac{\partial M}{\partial x}} ,
\label{W-Kubo-4}
\ee
where $M$ is the product of scattering matrices defined in
(\ref{M-def}). The above correlation function for the random
$SU(2)$ matrix $M$ can be related to the correlation function
(\ref{eta-KL}) and is equal to $8\,{\cal C}$. Therefore,
Eq.~(\ref{W-Kubo-4}) exactly coincides with its low-velocity
analogue (\ref{eta-KL}) and reproduces the result
(\ref{eta-high-v}).

\section{Discussion}
\label{section-discussion}

The main conclusion of this paper is that the Bardeen-Stephen
expression for the flux flow conductivity is extremely insensitive
to the details of the level correlations in the vortex core. We
calculated the energy dissipation rate for the case of the
Koulakov--Larkin two-comb level structure when the spectral
correlations are most pronounced. Such a statistics, classified as
the circular unitary ensemble of dimension two, is obtained for layered
superconductors within the region of the moderately clean limit
specified by Eq.~(\ref{upper-bound}). We found that the vortex
viscosity is the same for the regimes of discrete ($v<v_K$) and
continuum ($v>v_K$) spectra and coincides with the result
(\ref{eta-tau}) of the phenomenological $\tau$-approximation. The
viscosity $\eta$ determines the flux-flow dissipative conductivity
via the standard relation $\sigma_{xx}=ec\eta/\pi B$. Assuming a
cylindrical Fermi surface, we arrive at the result
(\ref{sigma-xx}) obtained previously via several quasiclassical
approaches
\cite{GK73,Bardeen-Sherman,LO76,Kopnin-Kravtsov,Kopnin94,Kopnin-Lopatin}.

Despite we found the same result for $\sigma_{xx}$ in the limits
of small and large vortex velocities, it is worth emphasizing that
the physics of energy dissipation is quite different in the two
limits. In the Kubo regime ($v>v_K$) the energy is pumped
continuously, whereas in the Landau-Zener regime ($v<v_K$) it is
absorbed in discrete portions of size $\omega_0$. The equivalence
between the $\tau$-approximation and the result of the exact
microscopic treatment is not very surprising in the Kubo regime,
as in this limit it is determined by the net sensitivity of the
spectrum to the vortex displacement rather than by interlevel
correlations. On the other hand, in the Landau-Zener regime, this
equivalence is a matter of coincidence: it relies on the fact that
the vortex Hamiltonian belongs to the circular {\em unitary}
universality class characterized by a level repulsion parameter
$\beta=2$.

The other outcome of the present work is that it provides a
justification for neglecting the Doppler-shift $T$-term,
Eq.~(\ref{T-kernel}), that was implicitly assumed in earlier
papers \cite{FS,LO98,KL}. Without the $T$-term, the system is
equivalent to a vortex at rest with impurities moving through its
core. The dissipation is then related to the change of the level
positions due to the motion of the impurities. On the other hand,
the $T$-term describes the electric field in the core, which is
the source of energy dissipation in the Bardeen-Stephen model ---
omitting this term is a bit confusing. Nevertheless, our analysis
indicates that the vortex viscosities calculated {\em with} and
{\em without} this term coincide in the moderately clean limit.
One may expect, however, that a correct treatment of this term is
crucial in the calculation of the Hall conductivity and in the
superclean limit.

In this paper we considered a strictly two-dimensional
superconductor (or layered superconductor with negligible
interlayer coupling). In the case of the Koulakov--Larkin circular
unitary ensemble statistics, the effect of interlayer hopping is
small provided that the effective quasiparticle temperature in the
core, $T_{\rm core}$, is smaller than $\omega_0(m_c/m_{ab})$,
where $m_c/m_{ab}$ is the effective mass anisotropy. This
condition ensures that the interlayer hopping amplitude is always
smaller than $\omega_0$ and can be neglected. This behavior is to
be contrasted with the one for the class $C$ random-matrix
statistics realized in the moderately clean case for weak
impurities ($\vartheta\ll1$) and in the dirty ($\Delta\tau\ll1$)
limit. In that case, the interlayer coupling leads to tunneling
from the $n$-th level in one layer to the $m$-th level (with $m
\neq n$) in the adjacent layer, thereby opening an interlayer
channel for energy dissipation. The competition between the
intralayer and interlayer channels may lead to a strong deviation
from the Bardeen--Stephen formula (\ref{sigma-xx}) and even to
hysteretic behavior of the current-voltage curve\cite{SF00}. On
the contrary, the two-comb spectrum is absolutely rigid: the
$n$-th level in one layer can have an avoided crossing only with
the $n$-th level in the adjacent layer and the tunneling between
these two levels does not result in dissipation.

\begin{acknowledgments}

We thank M.~V.~Feigelman for many useful discussions. This research
was supported by the SCOPES program of Switzerland, the Dutch
Organization for Fundamental Research (NWO), the Russian
Foundation for Basic Research under grant 01-02-17759, the program
``Quantum Macrophysics'' of the Russian Academy of Sciences, the
Russian Ministry of Science, the Russian Science Support
Foundation (M.~A.~S.), and the Swiss National Foundation.
M.~A.~S.\ thanks ETH Z\"urich for hospitality.

\end{acknowledgments}

\appendix

\onecolumngrid

\section*{Appendix}

Here, we calculate the off-diagonal correlation function $C_{\rm off-diag}$
defined in Eq.~(\ref{Coffdiag-def}). To this end we divide the angle
interval $[0,\pi]$ into many small pieces
$[\phi^{(k-1)},\phi^{(k)}]$ of width $\delta\phi \to 0$ so that
each piece contains one impurity at maximum. The transfer matrix
of the $k$-th interval is hence either
$M_{\phi^{(k)},\phi^{(k-1)}}=1$ (no impurities) or
$M_{\phi^{(k)},\phi^{(k-1)}}=M_i$ (if the angle of the $i$-th
impurity $\phi_i\in[\phi^{(k-1)},\phi^{(k)}]$). Then
\begin{align}
  C_{\rm off-diag} & {}
  = \frac 14 \sum_{k\neq p}
    \left<
      \tr M_{\pi,\phi^{(k)}}
        \frac{\partial M_{\phi^{(k)},\phi^{(k-1)}}}{\partial x}
        M_{\phi^{(k-1)},0}
    \times
      \tr M_{\pi,\phi^{(p)}}
        \frac{\partial M_{\phi^{(p)},\phi^{(p-1)}}}{\partial x}
        M_{\phi^{(p-1)},0}
    \right>
\nonumber \\ & {}
  = \frac 14 \sum_{k\neq p}
    \left<
      \tr R Y(k)
      \times
      \tr R M^\dagger_{\phi^{(p-1)},\phi^{(k)}}
        \tilde Y(p) M_{\phi^{(p-1)},\phi^{(k)}}
    \right>
\label{A1}
\end{align}
where $R = M_{\phi^{(k)},0} M_{\pi,\phi^{(k)}}$ and
\be
  Y(k) =
  \frac{\partial M_{\phi^{(k)},\phi^{(k-1)}}}{\partial x}
  M^\dagger_{\phi^{(k)},\phi^{(k-1)}},
\qquad
  \tilde Y(p) =
  M^\dagger_{\phi^{(p)},\phi^{(p-1)}}
  \frac{\partial M_{\phi^{(p)},\phi^{(p-1)}}}{\partial x} .
\ee
The representation (\ref{A1}) is suitable for averaging over
disorder, since one can independently average the matrices $R$,
$Y$, $\tilde Y$, and $M_{\phi^{(p-1)},\phi^{(k)}}$. The
statistical independence of $Y(k)$, $\tilde Y(p)$, and
$M_{\phi^{(p-1)},\phi^{(k)}}$ follows from the fact that the
intervals $[\phi^{(k-1)},\phi^{(k)}]$, $[\phi^{(k)},
\phi^{(p-1)}]$, and $[\phi^{(p-1)},\phi^{(p)}]$ do not overlap. As
we will see below, the correlator (\ref{A1}) is essentially
nonzero at $|\phi^{(k)}-\phi^{(p)}|\sim\phi_\tau$. The matrix $R$,
which couples $\phi^{(k)}$ and $\phi^{(k)}$ through the point
$\phi=\pi$, is the product of a large number of matrices $M_i$ and
looses all correlations within the interval
$[\phi^{(p-1)},\phi^{(p)}]$.

In calculating $\corr{Y(k)}$ and $\corr{\tilde Y(p)}$ with the
help of Eqs.~(\ref{M-i}) and (\ref{J-i}) only the fast phase of
$J_i$ should be differentiated,
\be
  \corr{Y(k)} = -\corr{\tilde Y(k)}
  = i \sigma_z 2k_F \cos\phi^{(k)} \,
    \delta\phi\int_0^\infty 2nr\,dr \frac{4|J(r)|^2}{(1+|J(r)|^2)^2}
  = i \sigma_z k_F \frac{\delta\phi}{\phi_\tau} \cos\phi^{(k)} ,
\ee
with the last relation following from the definition
(\ref{phitau}) of the angle coherence length $\phi_\tau\ll1$.
Taking the continuum limit $\delta\phi\to0$ we obtain
\be
  C_{\rm off-diag}
  = \frac{k_F^2}{4\phi_\tau^2}
    \int_0^\pi d\phi_1 d\phi_2 \cos\phi_1 \cos\phi_2
    \left<
      \tr R \sigma_z
      \times
      \tr R M^\dagger_{\phi_2,\phi_1} \sigma_z M_{\phi_2,\phi_1}
    \right> .
\ee
Averaging over $M_{\phi_1,\phi_2}$ according to
Eq.~(\ref{wf-decay}) and integrating over $(\phi_1+\phi_2)/2$ we
find
\be
  C_{\rm off-diag}
  = \frac{\pi k_F^2}{4\phi_\tau^2}
    \left< \tr R \sigma_z \times \tr R \sigma_z \right>
    \int_0^\infty d\phi \: e^{-\phi/\phi_\tau} \cos\phi ,
\ee
where the upper limit is substituted by infinity due to the fast
convergence of the integral. Finally, averaging over $R$ uniformly
distributed over the $SU(2)$ group and evaluating the remaining
integral, we arrive at Eq.~(\ref{Coffdiag}).

\twocolumngrid


\begin{thebibliography}{100}
\bibitem{e-ph}
The accurate calculation of the relaxation rate due to interaction
with phonons remains to be performed.
\bibitem{Wilkinson-comment}
Wlikinson considered only the orthogonal and unitary ensembles,
but the generalization to the symplectic ensemble is trivial.
\bibitem{KL-comment1}
The integral in Eq.~(21) of Ref.~\onlinecite{KL} should be normalized
by the area $2\pi^2$ of the 4-sphere.
\end{thebibliography}
\end{document}